\documentstyle[preprint,tighten,epsf,aps]{revtex}

\begin{document}
\def\be{\begin{equation}}
 \def \ee{\end{equation}}   
\def\bea{\begin{eqnarray}}\def\eea{\end{eqnarray}}
\preprint{NT@UW-01-05}
\title{Revealing 
Nuclear Pions Using 
 Electron Scattering}

\author{Gerald A. Miller}
                                
\address{Department of Physics\\
  University of Washington\\
  Seattle, Washington 98195-1560}

\maketitle

\begin{abstract}

A model for the pionic components of nuclear  wave functions is obtained from
light front dynamical calculations of binding energies and densities.
The  pionic effects are small enough to be
 consistent  with measured
nuclear di-muon production data and with  the nucleon sea. But the
pion effects  are large enough to
predict substantial nuclear
enhancement of the cross section for longitudinally polarized virtual
photons for
the kinematics accessible at Jefferson Laboratory.
\\shadow/pion/reveal.tex
\end{abstract}
\newpage
The belief that pions are the carrier of the nuclear force and hence
the dominant agent responsible for the binding of nuclei has been a basic
premise of nuclear physics since the time of Yukawa's original theory.
This tenet was strengthened via the 
observation of very significant effects of pion exchange currents in a variety
reactions involving electromagnetic probes of nuclei\cite{geb}.
Furthermore, chiral
Lagrangians involving pions and nucleons are believed to be the correct 
low-energy  representations of QCD\cite{weinberg}.

All of this basic physics was brought into question by 
lepton-nucleus deep inelastic scattering experiments and 
related measurements of  the production of Drell-Yan 
$\mu^+\mu^-$ pairs in   high energy proton-nucleus interactions.
  The lepton-nucleus deep inelastic scattering
observed by the 
EMC experiment \cite{emc,emcrevs} showed that there is a significant
difference between the parton distributions of free nucleons and
nucleons in a nucleus, including the substantial            result that
the nuclear structure function is supressed in the region
of the Bjorken variable 
$x\sim 0.5$. This means that
the valence quarks of bound nucleons carry less plus-momentum than
those of free nucleons. But partons must carry the total momentum
of the nucleus so the plus-momentum of non-nucleonic constituents must
be enhanced. One useful postulate was that it is nuclear pions 
which carry a larger fraction of the plus-momentum in
the nucleus than in free space\cite{chls,et}.
Such  models\cite{emcrevs} explain the shift
in the valence distribution and provide also 
an enhanced  nuclear  anti-quark distribution, 
observable in Drell-Yan
experiments \cite{dyth}. However, no such enhancement was 
observed\cite{dyexp}, and  no substantial  pionic enhancement 
was found in (p,n) reactions\cite{pn}. 
      These failures to observe the influence of
nuclear pions  were             termed    a severe crisis
for nuclear theory\cite{missing}.

But the
 magnitude of the crisis depends on the size of the pionic effects
expected from nuclear theories. The earliest calculations\cite{et}
included  enhancements  via the effects  of 
$\pi N\leftrightarrow\Delta $ transitions incorporated by summing
the RPA series. Another calculation,
involving significant $\Delta$ effects, evaluated
 the expectation value of the pionic number density operator
 in a variational
wave function\cite{bf} and the result  was characterized
as infinite nuclear matter  containing  
about 0.18 excess pions per nucleon. 
Both sets of calculations \cite{et,bf} led to significant nuclear pion content
and a corresponding enhancement of the nuclear anti-quark distribution,
which was not observed experimentally\cite{dyexp}. However, it was recognized
almost immediately 
 that one could reduce \cite{jm} the predicted sea enhancement 
by a very small increase           in the parameter          $g'$ 
used by \cite{et} to represent the
 short-ranged
nucleon-nucleon correlations. 
Thus it seems 
desirable to
make further calculations of nuclear wave functions in a manner which 
includes pionic components along with  nuclear saturation properties.

Furthermore, it is natural to expect that  calculations of 
high energy observables would benefit by using light front 
dynamics\cite{lcrevs}.
These considerations 
 led us to  use light front dynamics to 
compute nuclear wave functions\cite{miller00}.
One approach was to use 
 a new light-front one
  boson exchange potential\cite{rmgm98} to compute 
the saturation properties 
 of infinite nuclear matter in a Bruckner theory calculation which
obtained very reasonable values of the   binding energy
  density, and compressibility.
  This calculation goes beyond mean field theory, as is necessary to
obtain a non-zero pion content for nuclei with equal numbers of neutrons and
protons.
  Another result 
  is that the number of excess pions per nucleon is 0.05
which seems to be 
  in the  range shown\cite{jm}  as allowed  by the Drell-Yan data.
This  smaller value of the pion excess arises mainly    from the use of
one-boson exchange, in which the $\Delta$ is absent \cite{delta}.
Thus a  reasonable nuclear theory exists  which could       coexist with
the di-muon production data, but it is highly desirable to 
                            verify this set of dynamics by
 computing the di-muon production cross section and
 by observing a
 non-zero signal of pionic effects. These are  the aims of the present work.

One possibility for verification arises from measurements\cite{hermes}
of the ratio $R$,
of scattering of virtual photons in a longitudinal (L) or transverse
polarization state (T),
$R\equiv \sigma_{\rm L}/\sigma_{\rm T}$ from nuclei.
A  large value of $\sigma_{\rm L}$, and the corresponding violation 
of the Callan-Gross 
relation\cite{cagr},
 indicates the presence of nuclear bosons
as fundamental constituents of nuclei\cite{3m}. 
Indeed, the HERMES collaboration found a factor of five enhancement of
$R$, but this was for values of $x$ and $Q^2$ at  
 which the pionic effects are 
too small to be relevant\cite{3m}.

Here we                        study the possible
efffects of nuclear pions              on $\sigma_{\rm L}$ 
as a function
                 of $x$ and $Q^2$ and therefore to determine if such effects
are detectable.
We shall also compute those effects for the nucleon
target to predict its the pionic content. These effects are also
relevant to experiments aimed at determining the pion elastic form
factor \cite{garth}.

The start is to
confront  the deep inelastic scattering
data on the nucleon and the nuclear  
 di-muon production data which 
 place severe constraints on the
 pionic content of the nucleon\cite{tonymark}
 and nuclei, and ensure coexistence of the dynamical model 
with those data.  The necessary phenomenology
has been described often, see the reviews\cite{emcrevs}. The main point is that
the quark and anti-quark  distributions can be given as convolution of
the $q,\bar{q}$ distributions in a given nuclear hadronic constituent with
the light cone distribution functions: 
$f_{\pi/A}(y),f_{\pi/N}(y)$, the probability to find an excess
 pion in the nucleus (A) or nucleon (N),
with a plus-momentum given by $ym_N$. These quantities are
obtained from the
square of the ground state 
nuclear of nucleon  wave function, computed using light front dynamics, as: 
\bea
&& f_{\pi/N}(y)=m_N
\int d^2k_\perp \langle N\vert  n_\pi  ({\bf k}) \vert N\rangle
\\
&& f_{\pi/A}(y)={1\over A}m_N
\int d^2k_\perp \langle A\vert n_\pi  ({\bf k}) \vert A\rangle-f_{\pi/N}(y)
\label{def}
\eea
where $ n_\pi({\bf k}=({\bf  k}_\perp, k^+))$ is the  pion number operator.
The function $f_{\pi/N}(y)$ is 
constrained by the data on the nucleon sea which restrict the 
$\bar u$ and $\bar d$ distributions to be similar\cite{tonymark}.
The important input  is the $\pi N$ form factor, which we take to be of the
Cloudy Bag Model form \cite{cbm} with a bag radius,  0.9 fm, large enough
to be consistent with the constraints.

\begin{figure}
\unitlength1cm
\begin{picture}(10,8)(1,-8.5)
  \includegraphics{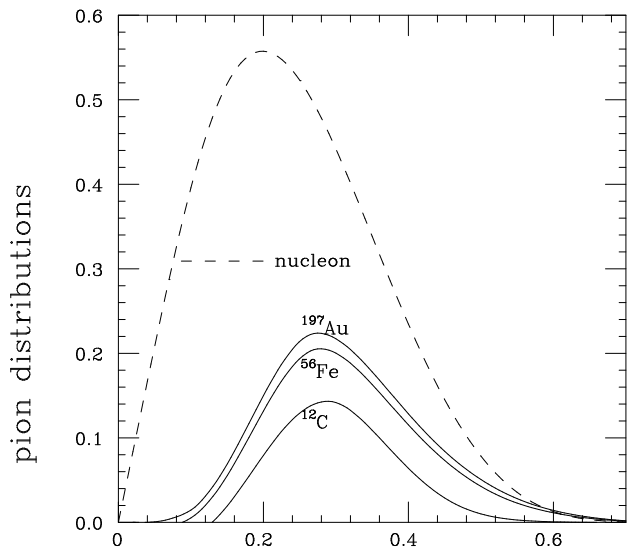}
\end{picture}
\caption{Excess pion distribution functions for nuclei (solid) and
  for the nucleon (dashed).
  The results for different nuclei $^{12}C,^{56}Fe$,
 and $^{197}Au$ are obtained from the Fermi gas model with values of
  $k_f=1.2,1.3$ and 1.34 fm$^{-1}$[22].}
  \label{fig:pidens} 
\end{figure}

The quantity   $f_{\pi/A   }(y  )$ controls the nuclear 
pionic effects. We would like to obtain this
directly from our latest light front calculation\cite{rmgm98}, of nuclear
matter properties. This calculation
 provides  the integral quantity for the excess number of
pions as
$ \int_0^\infty \;dy\; f_{\pi/A}(y)=0.05,$ 
but not the function $f_{\pi/A}$        itself. 
However in all models this function is expected to be vanish for 
very large and 
very small values of $y$ and smoothly rise to a maximum, at about 
$y=0.2,0.3$. Thus the overall strength of the effects
in Drell-Yan scattering and $\sigma_{\rm L}$ we wish to examine here
are not expected to be very sensitive to details or have much model dependence
as long as the overall normalization is constrained.   
Therefore  we revive     our
old calculation \cite{jm}, but now using  
parameters which yield 0.05 for            the      number of excess pions
in nuclear matter.
The simplest way to achieve this value is to vary  the value of $g'$. The
relation
 $g'=0.795$  gives the desired value for nuclear matter and  also 
 that excess nuclear pions carry  a fraction  0.016  of the nuclear
 plus momentum. We account for the difference between $Fe$ and nuclear
 matter by reducing the Fermi momentum from 1.37 fm$^{-1}$ to  1.3
 fm$^{-1}$\cite{kf}.
 This reduces the excess pion number to
 0.04 and the momentum fraction to 0.135.
 The results for the pionic distribution functions are shown in
Fig.~\ref{fig:pidens}.
We see that $f_{\pi/N,A}$ differ from zero  for $y\sim 0.2-0.5$, but
$f_{\pi/A}$ vanishes rapidly as $y$ approaches 0. This means that pionic
effects are irrelevant to understanding the HERMES effect, which occurs
for $x_{Bj}\sim 0.01$.
\begin{figure}
\unitlength1cm
\begin{picture}(10,8)(1,-8.5)
  \includegraphics{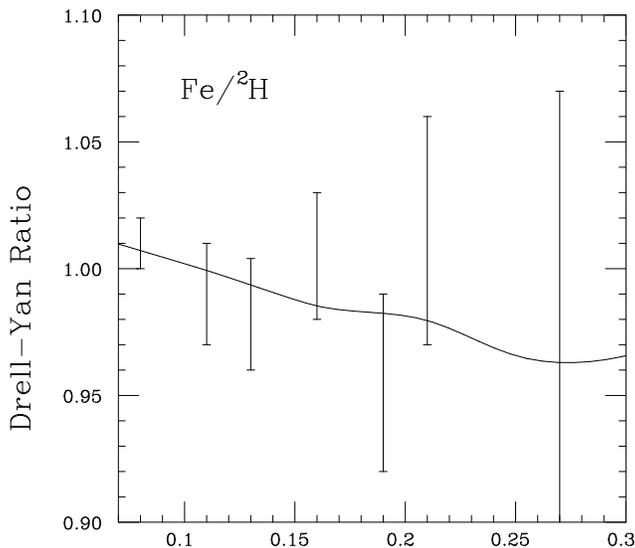}
\end{picture}
\caption{Ratios of the Drell-Yan di-muon ratio per nucleon, for positive
  $x_F=x_p-x_t$. The data are from Ref.~[8]. }
\label{fig:drell}
\end{figure}
Using $f_{\pi/A}$  one can compute the
contribution
 to the 
 di-muon production cross section in which a quark from the incident
 proton
 is annihilated by a nuclear anti-quark (from either a nucleon or a pion)
 to form a virtual photon which
 emerges as a di-muon--$\mu^+\mu^-$ pair.
The pionic contributions must be added to the contributions of the nucleons.
The nucleonic effects  are computed
using  the formalism of \cite{Jung:1988jw} and \cite{Dieperink:1991mw}.
 The key parameter is the  average separation energy, which is
take as 56 MeV, as obtained from an extrapolation of the nuclear matter
value
of 71 MeV.
Such parameters allow 
a reasonable qualitative
representation of the nuclear deep inelastic scattering data
\cite{emcrevs,Dieperink:1991mw,ex1}.

With the present  model, 
it 
is the nucleonic effects 
which  control the computed Drell-Yan results.
These are shown in Fig.~\ref{fig:drell}, for values
of the target plus momentum fraction, $x_t$ large enough so that     
 the effects of shadowing
are absent. The compuations are made using the experimental
acceptance\cite{dyexp}.
We see that it is possible to reproduce the experimental data  with a
 theory which is based on detailed nuclear dynamics such as light front
 Bruckner theory \cite{rmgm98}.


The question remaining is, ``Can there
be a signal to observe these pions?'' Thus we turn to  the role of
pionic effects on
$\sigma_L$ for nuclear and nucleon targets. 
The pionic contribution to the   
hadronic tensor is given by:
 \be
\delta^\pi W^{\mu\nu}={1\over 4\pi M_A}\int d^4\xi\; e^{iq\cdot\xi}\;\langle
P|J_\pi^\mu(\xi)J_\pi^\nu(0)|P\rangle, 
\label{start}\ee
where $\vert P\rangle$ represents the nucleon or nuclear target ground state
of total momentum $P$, and
the momentum of the virtual photon is
$q=(\nu,{\bf 0}_\perp,-\sqrt{\nu^2+Q^2})$.
The current operator which accounts for the effects arising from the
nuclear pions  is $J_\pi^\mu(\xi)$:
\be J_\pi^\mu=i(\phi^*\partial^\mu \phi-\phi \partial^\mu \phi^*),\ee
with  $\phi$ as the complex pion
field operator.
The effects of the pion charge form factor $F_\pi(Q^2)$ are    
included below.

We are concerned with $\sigma_{\rm L}$, and use
the standard general formula\cite{roberts,halzen}; 
\bea
\sigma_L={Q^2 4\pi^2\alpha\over (Q^2+\nu^2)^{3/2}}\;W^{00}.\label{stand}\eea
We now  evaluate $\delta^\pi W^{00}$  by using             $J_\pi^{0}$       
in
Eq.~(\ref{start}). Then insert a set of states which asymptotically contain
a  nucleus and a free pion of momentum $p^\mu$. The sum over nuclear
states can be performed using closure if one uses light front variables
($q^\pm=q^0\pm q^3$) to
describe the momentum of the photon and of the nuclear pion
$({\bf k}_\perp,k^+)$. 
For large enough values of $\nu$ the interactions of
the struck pion with the residual nucleus may be  ignored, 
 the  energy of the
outgoing pion may be  taken as $\nu$, and
the effects of non-zero values of nuclear
excitations energies (proportional to $q^+/q^-$) appearing in the
energy-momentum conserving delta function inherent in  Eq.~(\ref{start})
may be ignored.
Then the  result of a straightforward
evaluation leads to the result:
\bea
\delta^\pi
W^{00}(N,A)={\nu^2\over 2Q^2} {2\over 3}{f_{\pi/(N,A)}(\xi)\over M_N}
F_\pi^2(Q^2),
\label{key} \eea
where
$F_\pi(Q^2)$ is taken as a monopole form consistent
with the observed mean square radius, and 
the factor of 2/3 accounts for the fact that only charged 
pions enter in electron scattering (with N=Z). Note that energy
 momentum conservation  gives the relation $ k^+=M_N \xi={Q^2\over q^-},$   
and          $\xi$ is the Nachtmann variable. 

\begin{figure}
\unitlength1cm
\begin{picture}(10,8)(1,-8.5)
  \includegraphics{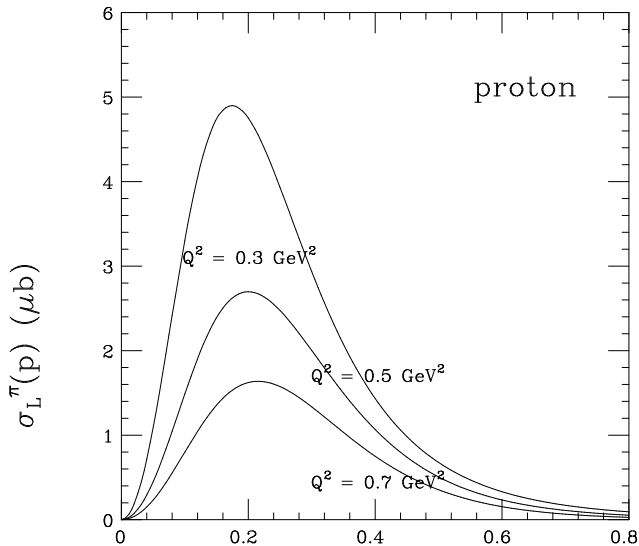}
\end{picture}
\caption{Pionic contribution to the photon longitudinal cross section on the
  proton, as a function of $Q^2$ and
  $x$. }
\label{fig:piprot}
\end{figure}

The pionic contribution to the
longitudinal cross section for a proton (p=N) target, $\sigma_L^\pi({\rm p})$.
 is obtained from Eqs.~(\ref{stand}) and (\ref{key}), 
and is displayed in Fig~(\ref{fig:piprot}). Cross sections of this size are
routinely observable. The decrease with increase of $Q^2$ is caused by
the pion form factor. The value of   $\sigma_L^\pi({\rm p})$ for
$x=0.2,\;Q^2=0.7\; {\rm GeV}^2$ corresponds to a contribution to $R$ of the
proton of about 0.04,  small enough to avoid any contradiction with
existing data.

We now turn to the case of nuclear targets, 
The standard general formula (\ref{stand}) 
allows us to relate $\delta^\pi W^{00}$ to the corresponding
contribution (per nucleon) $\delta^\pi\sigma_{\rm L}$ to the longitudinal
cross section,
so that  the nuclear longitudinal cross section  (per nucleon)
is given by
$
\sigma_L(A)=\sigma_L(D)+\delta^\pi \sigma_L.$ 
It is desirable to present a ratio
${\sigma_L(A)/ \sigma_L(D)}$, or equivalently
${\sigma_L(A)/\sigma_T(D)R_D}$,    
which allows the use of 
   parameterizations for  $F_2(D)$ and $R(D)$. Thus we obtain
the result:
\be
{\sigma_L(A)\over \sigma_L(D)}=1+{Q^4\over (Q^2+\nu^2) \nu}
{\delta^\pi W^{00}\over A\; F_2^D R_D}
(1+R_D), \label{gen}\ee
which may be  evaluated
using Eq.~(\ref{key}) to be: 
\be
{
\sigma_L(A)\over \sigma_L(D)}=1+x{2\over3}f_\pi(\xi)
{\nu^2\over (Q^2+\nu^2) }
{F_\pi^2(Q^2)\over F_2^D R_D}
(1+R_D), \label{spec}\ee
where as usual
$x=Q^2/2M_N\nu$. Note that only the pionic effects are included here.
The effects of nuclear vector and scalar mesons, so important in 
understanding the HERMES effect\cite{3m}, are not included because these
appear at much lower values of $x$ than we shall consider.

\begin{figure}
\unitlength1cm
\begin{picture}(10,8)(1,-8.5)
  \includegraphics{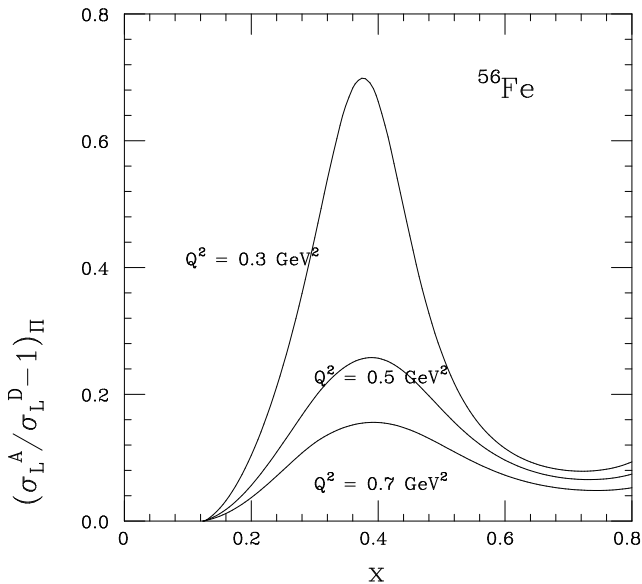}
\end{picture}
\caption{Enhancement of longitudinal cross sections, as a function of $Q^2$ and
  $x$. }
\label{fig:sigl}
\end{figure}

The results of our calculation of $\sigma_{\rm L}$ on $^{56}Fe$
are shown in Fig.~\ref{fig:sigl} which shows the
difference between the left hand side of Eq.~(\ref{spec}) and unity.
Results for other nuclei can be obtained
by using the different pion distribution functions of Fig.~\ref{fig:pidens}.

Large effects are predicted for a quantity which is readily
measurable at Jefferson Laboratory.
The largest effects obtained with the
 lowest value of  $Q^2=0.3\; {\rm GeV}^2$, 
correspond to very low values of $\nu\sim 0.4\; {\rm GeV}$, where
there may be substantial corrections to the light-front
closure approximation 
because
$q^+$ is not vanishingly small. While  there seems to be
no reason why the effects of such corrections would cancel the effects shown 
here, the interpretation of  the data would be simplified by observing effects
at larger values of $Q^2$. In this case one obtains smaller values of $q^+$,
 and 
  larger values of $\nu$, but  the   outgoing pion 
energies are in the vicinity of the $\pi$-nucleon resonance region, so that 
final state interaction effects  could change the shape 
of the curve. However, since
the variable $\sigma_L$ involves an inclusive measurement the overall strength
as represented by, e.g. the area under the curve, should not be affected very 
much.

The  substantial 15-30\% effects,
for values of $x\sim0.2-0.3$,  present an 
 opportunity to observe a significant nuclear enhancement of $\sigma_L$. This
 is  an excellent opportunity to
unravel a significant long-standing 
mystery  involving the absence of nuclear pionic effects.

\section*{Acknowledgments}
This work is partially supported by the USDOE. I thank H.~Blok, 
S.J.~Brodsky and M. Strikman for useful discussions.


\begin{references}

\bibitem{geb} Nucleon-Nucleon Interaction'', G.E. Brown and A.D. Jackson 
North-Holland (Amsterdam) 1976; ``Pions and Nuclei'', T. Ericson and W. Weise
Oxford Science Publications, (New York) 1987.
\bibitem{weinberg}
S.~Weinberg,
Physica A {\bf 96}, 327 (1979).
\bibitem{emc} J. Aubert {\it et al.}, Phys.\ Lett.\ 
{\bf 123B}  (1982) 275; 
R.G. Arnold {\it et al.,} Phys. Rev. Lett. {\bf 52} (1984) 727; 
A. Bodek {\it et al.}, Phys. Rev. Lett. {\bf 51}  (1983) 534. 

\bibitem{emcrevs} G.~Piller and W.~Weise,
Phys.\ Rept.\  {\bf 330}, 1 (2000);
 L.L. Frankfurt and M.I. Strikman, Phys. Rep. {\bf 160}(1988)  235; 
M. Arneodo, Phys. Rep. {\bf 240} (1994) 301; 
D.F. Geesaman,
K. Saito, A.W. Thomas, Ann. Rev. Nucl. Part. Sci. {\bf 45} (1995) 337.
         

\bibitem{chls} C.H.~Llewellyn Smith {\bf B128} (1983) 107.
\bibitem{et}  M.~Ericson and A.W.~Thomas, Phys. Lett. {\bf B128} (1983) 112. 
\bibitem{dyth} R.P. Bickerstaff, M.C. Birse, and G.A. Miller,
Phys. Rev. Lett. {\bf 53}, (1984) 2532; 
M. Ericson and A.W. Thomas,
Phys. Lett. {\bf 148B} (1984) 191. 
\bibitem{dyexp} D.M. Alde {\it et al.},
  Phys. Rev. Lett. {\bf 64} (1990) 2479. 
\bibitem{pn} 
  L. B. Rees et al., Phys. Rev. C 34, 627 (1986); 
    J. B. McClelland et al., Phys. Rev. Lett. 69, 582 (1992); 
   T. N. Taddeucci et al., Phys. Rev. Lett. 73, 3516 (1994). 

\bibitem{missing} G.F. Bertsch, L. Frankfurt, and M. Strikman,
Science {\bf 259} (1993) 773. 

\bibitem{bf} B. Friman, V.R. Pandharipande and R.B. Wiringa, Phys. Rev. Lett
  {\bf 51} (1983) 763. 
\bibitem{jm} H. Jung and G.A. Miller, Phys. Rev. {\bf C41}(1990) 659. 

\bibitem{lcrevs} S. J. Brodsky, H.C.~Pauli,
S.S.~Pinsky, Phys. Rep. {\bf 301}, 299-486 (1998);
 L.L.~Frankfurt and M.I.~Strikman,
Phys. Rep. {\bf 76},  (1981) 215. 
\bibitem{miller00}
G.~A.~Miller,
Prog.\ Part.\ Nucl.\ Phys.\  {\bf 45}, 83 (2000).
\bibitem{rmgm98}G.A. Miller, Phys. Rev. C 
{\bf 56} (1997) 2789;
G.A.~Miller and R.~Machleidt,
Phys. Lett. {\bf B455} (1999) 19; 
Phys. Rev. {\bf C60}  (1999) 035202-1. 
\bibitem{delta}
This number of excess  pions is far smaller
than the $\sim0.2$  obtained  earlier\cite{bf}. Much of 
the difference 
is caused by    the absence of $\Delta$ effects in our one-boson-exchange
formulation and the presence of such, computed with a hard form factor
in Ref.~\cite{bf}. Recent electron-scattering data V.~V.~Frolov {\it et al.},
Phys.\ Rev.\ Lett.\ {\bf 82}, 45 (1999)
show    that the electromagnetic $N\Delta$ transition form
factor falls rapidly with increasing momentum transfer, so that it could be
more
appropriate to use a soft form factor. Thus including no nuclear $\Delta$'s
may be a better approximation than including huge amounts.
\bibitem{hermes} 
K. Ackerstaff et al,
Phys.\ Lett.\ B {\bf 475}, 386 (2000)
\bibitem{cagr} C.G.~Callan and D.J.~Gross, Phys. Rev. Lett. {\bf 22}
  (1969) 22.
\bibitem{3m} G.A.~Miller, S.J.~Brodsky and M. Karliner,
Phys.\ Lett.\  {\bf B481}, 245 (2000).

\bibitem{garth}
J.~Volmer {\it et al.}, 
Phys.\ Rev.\ Lett.\ {\bf 86}, 1713 (2001).
\bibitem{tonymark}
A.~W.~Thomas,
Phys.\ Lett.\ B {\bf 126}, 97 (1983).
L.~L.~Frankfurt, M.~I.~Strikman, L.~Mankiewicz, A.~Schafer, E.~Rondio, A.~Sandacz and V.~Papavassiliou,
Phys.\ Lett.\ B {\bf 230}, 141 (1989).
\bibitem{cbm} S.\ Th\'eberge, A.\ W.\ Thomas and G.\ A.\ Miller,
Phys. Rev. {\bf D22} (1980) 2838; {\bf D23} (1981) 2106(E). A.\ W.\ Thomas,
S.\ Th\'eberge, and G.\ A.\ Miller, Phys. Rev. {\bf D24} (1981) 216. 

\bibitem{kf}
 E.~J.~Moniz, I.~Sick, R.~R.~Whitney, J.~R.~Ficenec, R.~D.~Kephart and W.~P.~Trower,
Phys.\ Rev.\ Lett.\ {\bf 26}, 445 (1971).
  










\bibitem{Jung:1988jw}
H.~Jung and G.~A.~Miller,
Phys.\ Lett.\ B {\bf 200}, 351 (1988).

\bibitem{Dieperink:1991mw}
A.~E.~Dieperink and G.~A.~Miller,
Phys.\ Rev.\ C {\bf 44}, 866 (1991).
\bibitem{ex1} Previous work required momentum 
missing by nucleons to be carried only by excess
nuclear pions, and this condition
yields the momentum sum rule.  
However, vector mesons can carry a significant fraction of the nuclear plus
momentum:
G.A. Miller, Phys. Rev. C {\bf 56} (1997) R8, so that
the momentum sum rule need not be satisfied
by nucleons and pions.  

\bibitem{roberts}
R.~G.~Roberts,
{\it  Cambridge, UK: Univ. Pr. (1990) 182 p. (Cambridge monographs on mathematical physics)}.
\bibitem{halzen}F.~Halzen and A.~D.~Martin,
{\it  New York, Usa: Wiley ( 1984) 396p}.


\bibitem{Abe} 
K.~Abe {\it et al.}  [E143 Collaboration],
Phys.\ Lett.\  {\bf B452}, 194 (1999)
[hep-ex/9808028].
\bibitem{f2ref} 
M.~Arneodo {\it et al.}  [New Muon Collaboration.],
Phys.\ Lett.\ {\bf B364}, 107 (1995)
[hep-ph/9509406].





\bibitem{Whitlow:1990gk}
L.~W.~Whitlow, S.~Rock, A.~Bodek, E.~M.~Riordan and S.~Dasu,
Phys.\ Lett.\ {\bf B250}, 193 (1990).





\end{references}
\end{document}